\documentclass[reprint,aps,pra,superscriptaddress]{revtex4-2}
\usepackage{graphicx}
\usepackage{amsmath}
\usepackage{bbm}
\usepackage{amssymb}
\usepackage{amsthm}
\usepackage{amsfonts}
\usepackage{graphicx}
\usepackage{xcolor}
\usepackage{hyperref}
\usepackage{hyperref}

\hypersetup{
	colorlinks = true,
	citecolor = red,
	urlcolor = blue,
	linkcolor=blue
}

\newcommand{\eqnref}[1]{Eq.~(\ref{#1})}
\newcommand{\figref}[1]{Fig.~\ref{#1}}
\newcommand{\sfigref}[2]{Fig.~\hyperref[#1]{\ref{#1}#2}}

\newcommand{\secref}[1]{Sec.~\ref{#1}}

\newcommand{\paraheading}[1]{ \emph{#1}---}

\newcommand{\SMRef}[1]{{see \cite{SM}, \protect{#1}}}

\newcommand{\expect}[1]{\left\langle #1 \right\rangle}
\newcommand{\bra}[1]{{{\langle #1 |}}}
\newcommand{\ket}[1]{{| #1 \rangle}}

\newcommand{\expectS}[3]{\langle #1 | #2 | #3 \rangle}
\newcommand{\ci}{\mathrm{i}}

\newcommand{\inv}[1]{{#1}^{-1}}
\newcommand{\gcenter}[1]{\mathcal{Z}\left(#1\right)}
\newcommand{\cyclicg}[1]{\mathbb{Z}_{#1}}

\newcommand{\schurmult}[1]{H^2\left(#1, \text{U}(1)\right)}
\newcommand{\secondcohomology}[2]{H^2\left(#1,#2\right)}
\newcommand{\commutatorsg}[1]{[#1,#1]}

\newcommand{\tr}[1]{\text{tr}#1}

\newcommand{\biggroup}{\mathbbm{G}}
\newcommand{\elembigg}[1]{\mathbbm{#1}}

\newcommand{\mytitle}{Lieb-Schultz-Mattis theorem from gauge constraints}
\newcommand{\authorOne}{Bhandaru Phani Parasar}

\begin{document}
	
	\title{\mytitle}
	
	\author{\authorOne}
	\email{bhandarup@iisc.ac.in}
	\affiliation{Centre for Condensed Matter Theory, Department of Physics, Indian Institute of Science, Bangalore 560012, India}
	
	\begin{abstract}
We construct a $\mathbb{Z}_2 \times \mathbb{Z}_2$ gauge theory coupled to matter on a one-dimensional chain, aiming to study the ground-state physics in the Gauss law subspace. We show that the theory in the Gauss law subspace has a U$(1)$ symmetry whose generator commutes with lattice translations, but anticommutes with the lattice reflection operator. This leads to a Lieb-Schultz-Mattis (LSM) theorem that \emph{always} rules out a trivial gapped ground state in the Gauss law subspace, if the hamiltonian is invariant under translations and reflection. Any point in the parameter space must realize either a spontaneously symmetry broken (SSB) ground state, or a gapless ground state. Imposing the Gauss law is pivotal for the existence of the U$(1)$ symmetry, and hence of the LSM theorem. We thus demonstrate a novel mechanism to obtain an LSM-type theorem, wherein the symmetry responsible for the theorem originates from the kinematic constraints of a gauge theory. We identify a point in the parameter space at which the system is gapless. At the gapless point, the excitations admit a description in terms of free Dirac fermions with a constraint on the total fermion number. The asymptotic behavior of the two-point correlation function of the simplest local gauge-invariant quantity at the gapless point is found to be  $ \propto \cos{(\pi r)}\,r^{-2/9}$, where $r$ is the lattice separation between the two points. This model is also a natural platform to study phase diagram topological defects residing in families of SSB phases.
	\end{abstract}
	
	\maketitle

	\paraheading{Introduction}
	Lieb-Schultz-Mattis (LSM)\cite{LSM1961}-type theorems place nontrivial constraints on the ground-state properties of some quantum many-body systems with spatial and internal symmetries. These theorems can constrain the possible low-energy physics based on the manner in which symmetries act on the microscopic degrees of freedom. Originally formulated for the spin-$1/2$ antiferromagnetic Heisenberg chain with SO$(3)$ symmetry, the theorem has been extended \cite{Affleck_Lieb1986, Oshikawa_Affleck1997,Oshikawa2000,Hastings2004,Chen_Wen2011,Parameswaran_Vishwanath2013,Watanabe_Zaletel2015,Watanabe_PRB2018,Tasaki2018,Ogata_Tasaki2021,Kobayashi_Ryu2019,Else_Thorngren2020,Dubinkin_Hughes2021,Burnell_Prem2024,Yao_Furusaki2022,Yao_Hsieh2024} to higher-dimensional systems, and systems with general internal and spatial symmetries. 
	
	When applicable, an LSM-type theorem prohibits a trivial gapped ground state for the system: the ground state must spontaneously break a symmetry under consideration, or be gapless, or have  a nontrivial degeneracy on the torus. Recent studies reveal a compelling link between LSM-type theorems and quantum anomalies \cite{Cheng_Bonderson2016,Cho_Ryu2017,Jian_Xu2018,Else_Thorngren2020,Cheng_Seiberg2023,Seifnashri2024,Aksoy_Tiwari2024,Ebisu_Cao2026}. Considering the formidable challenge of resolving ground-state properties of a general quantum many-body system, the utility of LSM-type theorems is paramount.

	There are key questions to be explored concerning the general possibility of LSM-type theorems. Ref.~\cite{Kobayashi_Ryu2019} proves an LSM theorem in the context of a pure U$(1)$ gauge theory. Is it possible to obtain an LSM-type theorem in a theory of matter coupled to gauge fields with a discrete gauge group? In particular, can the kinematic structure of a gauge theory lead to nontrivial constraints on the low-energy properties of the theory in a Gauss law subspace?
	
	Gauge theories \cite{Kogut1979} play an indispensable role in condensed matter physics. Gauging a global symmetry of ``matter" degrees of freedom involves promoting the global symmetry to a local one by introducing auxiliary ``gauge" degrees of freedom and imposing a Gauss law constraint.  The study of many strongly correlated systems \cite{Senthil_Fisher2000,Moessner_Fradkin2001,Wen2002} leads to a description in terms of a theory of matter coupled to gauge fields. Gauge theories also provide a natural setting to describe and explore novel phases and critical phenomena \cite{Senthil2306.12638,Nandkishore_Senthil2012,Banerjee_Wiese2014,Gazit_Vishwanath2017,Gazit_Wang2018,Borla_Moroz2022,Das_Shenoy2025}.
	
	In this paper, we answer the above-raised questions in the affirmative by demonstrating how an LSM-type theorem arises in the Gauss law subspace of a $\cyclicg{2} \times \cyclicg{2}$ gauge theory in one spatial dimension. We construct a $\cyclicg{2} \times \cyclicg{2}$  gauge theory coupled to matter on a one-dimensional chain hosting a three-dimensional Hilbert space on each site and link. We show that the theory in the Gauss law subspace has a U$(1)$ symmetry, whose generator commutes with the lattice translation operator, but \emph{anticommutes} with the lattice reflection operator. This gives rise to an LSM theorem for the theory in the Gauss law subspace. At all values of the parameters, the theory is prohibited to realize a trivial gapped ground state. The origin of the U$(1)$ symmetry, and hence the LSM theorem, is in imposing the Gauss law. We thus uncover a novel mechanism to realize an LSM-type theorem, where the symmetry  responsible for the theorem originates from the kinematic structure of a gauge theory (Gauss law).
	We identify a gapless point in the parameter space with the aid of a nonlocal mapping to fermionic degrees of freedom. At the gapless point, the excitations are described by free Dirac fermions. We also study the asymptotic behavior of correlation functions at the gapless point using results from the theory of Toeplitz determinants.

	\paraheading{Model}
	Consider a one-dimensional periodic chain that hosts a three-dimensional Hilbert space on each site and link of the chain.  Let $L$ be the length of the chain ($L$ sites and $L$ links), where $L$ is even. Let $\mathcal{V}$ denote the combined Hilbert space of the site (``matter") and the link (``gauge") degrees of freedom. Then, $\text{dim}(\mathcal{V}) = 3^{2L}$.  With $\sigma$ and $\tau$ labeling the sites and the links of the chain respectively, we adopt a numbering convention in which link $j \tau$ joins the sites $j \sigma$ and $j+1 \,  \sigma$ (see \figref{fig:model_mu}(a)). Now, we define the hamiltonian for the combined system of matter and gauge degrees of freedom to be
	\begin{equation}\label{eqn:ham}
		H= \sum_{j, \alpha=x,y,z}  \left( t_\alpha S^\alpha_{j \sigma} S^\alpha_{j \tau} S^\alpha_{j+1 \sigma} - K_\alpha\left(S^\alpha_{j \tau}\right)^2 \right)
	\end{equation}
	Here, $S^\alpha_{j \sigma}$, for $\alpha=x,y,z$ ($S^\alpha_{j \tau}$) on each site $j \sigma$  (link $j \tau$) are $3\times 3$ spin-$1$ operators satisfying $[S^\alpha_{j \sigma},S^\beta_{j' \sigma}] = \ci \delta_{jj'} \sum_\gamma  \epsilon^{\alpha \beta \gamma} S^\gamma_{j \sigma}$ (A similar algebra is satisfied by the operators $S^\alpha_{j \tau}$), and $[S^\alpha_{j \sigma}, S^\beta_{j' \tau}]=0$. The parameters $t_\alpha,K_\alpha$ are real numbers~\footnote{Without loss of generality, $K_z$ may be set to zero because of the identity $\sum_\alpha (S^\alpha)^2 =2$.}. To verify that this is a gauge theory, we consider the operators $\Sigma^\alpha_{j \sigma } = \exp{(\ci \pi S^\alpha_{j \sigma})}= 1-2 (S^\alpha_{j\sigma})^2$ on each site $j \sigma$. The operators $\Sigma^\alpha_{j \tau}$ on each link $j \tau$ are analogously defined. Now, associated with a site $j \sigma$, we define the local operators $A^\alpha_j = \Sigma^\alpha_{j-1 \tau} \Sigma^\alpha_{j \sigma} \Sigma^\alpha_{j\tau}$. It can be verified that $[H,A^\alpha_j]=0$, and that the operators $A^\alpha_j$ satisfy the algebra in \eqnref{eqn:gt_algebra}. Hence, we conclude that the hamiltonian in \eqnref{eqn:ham} represents a  $\cyclicg{2} \times \cyclicg{2}$ gauge theory. 
	\begin{equation}\label{eqn:gt_algebra}
		[A^\alpha_j, A^\beta_{j'}]=0, \,\,
		(A^\alpha_j)^2 = A^x_j A^y_j A^z_j = 1
	\end{equation}
	 We note that this construction is equivalent to a gauging of the $\cyclicg{2} \times \cyclicg{2}$ global symmetry of the spin-$1$ XYZ chain using a \emph{three-dimensional} representation of $\cyclicg{2} \times \cyclicg{2}$ ($1,\Sigma^\alpha_{j \tau}$) that does not contain the trivial representation, on each link $j \tau$. This is to be contrasted with the  standard formulation \cite{Haegeman_Verstraete2015,Blanik_Schuch2025} of using the regular representation (which is four-dimensional in this case) on the links. The Kennedy-Tasaki transformation \cite{Kennedy_Tasaki1992}, which maps a symmetry protected topological phase to a spontaneously symmetry broken phase, has been interpreted  \cite{Li_Zheng2023} as twisted gauging \cite{Blanik_Schuch2025} for the spin-$1$ XYZ chain. Our construction described above corresponds to an ordinary gauging of the spin-$1$ XYZ chain.
	
	Let us discuss spatial symmetries of the model. Let $L$ be even, $T$ be the lattice translation operator (that translates by one unit cell), and let $R$ be the operator implementing the reflection of the periodic chain about a line passing through the sites $1,1+L/2$. Then, we have $T^L=R^2=1$, $R T \inv{R} = \inv{T}$, and
	\begin{equation}\label{eqn:TandR}
		\begin{split}
			T S^\alpha_{j \sigma} \inv{T} &= 	 S^\alpha_{j+1 \sigma} , \,\,\, 	T S^\alpha_{j \tau} \inv{T} = 	 S^\alpha_{j+1 \tau} \\
			R S^\alpha_{j \sigma} \inv{R} &= S^\alpha_{L+2-j \sigma}, \,\,\, 	R S^\alpha_{j \tau} \inv{R} = S^\alpha_{L+1-j \tau}
		\end{split}
	\end{equation}
	Since the coefficients $t_\alpha,K_\alpha$ are uniform, the hamiltonian in \eqnref{eqn:ham} is invariant under lattice translations and reflections. i.e.,~$[H,T]=[H,R]=0$.
	 
	For the local Hilbert space on each site and link, we work with a basis $\{\ket{\alpha}\}$, $\alpha=x,y,z$ in which the operators $S^\alpha$ act as $S^\alpha\ket{\beta}   = \ci \sum_{\gamma}  \epsilon^{\alpha \beta \gamma} \ket{\gamma}$. This clearly satisfies the spin algebra. In this representation, $S^\alpha$ have the matrix form given in \eqnref{eqn:Smat} and $(S^x)^2,(S^y)^2,(S^z)^2$ are all simultaneously diagonal: $P^\alpha := 1-  (S^\alpha)^2$ is the projector onto the state $\ket{\alpha}$. i.e.,~$P^\alpha \ket{\beta} = \delta^{\alpha \beta}\ket{\beta}$ (no sum over $\beta$).
	\begin{equation}\label{eqn:Smat}
	S^x=\begin{pmatrix}
		0 & 0 & 0\\0 & 0&-\ci \\ 0 & \ci &0
	\end{pmatrix}, S^y=\begin{pmatrix}
		0 & 0 & \ci \\
		0 & 0&0 \\ -\ci &0&0
	\end{pmatrix}, S^z=\begin{pmatrix}
		0 &-\ci&0 \\ \ci&0&0\\0&0&0
	\end{pmatrix}
	\end{equation}
	The tensor product basis  constructed using this local basis at each site and link is $\{\otimes_j \ket{\alpha_j}_{j \sigma} \ket{\beta_j}_{j \tau}, \alpha_j,\beta_j \in \{x,y,z\} \}$. The first term in \eqnref{eqn:ham}, $h^\alpha_j =S^\alpha_{j \sigma} S^\alpha_{j \tau} S^\alpha_{j+1 \sigma}  $ has vanishing diagonal elements in this tensor product basis, while the second term is diagonal. Hence we call them the ``hopping term" and the ``potential term" respectively. 

	Before we proceed further, it will be useful to introduce the operators $\mathcal{Z}, \mathcal{X}$ acting on the local three-dimensional Hilbert space of each site and link. We want to label the basis states $ \{\ket{\alpha}, \alpha=x,y,z \}$ with the distinct eigenvalues of $\mathcal{Z}$, and $\mathcal{X}$ to be a ``cyclic raising operator" on this basis. We define
	\begin{equation}\label{eqn:calZcalX}
		\begin{split}
		\mathcal{Z} &:=
		-\sum_{\alpha}{\xi_\alpha} \left(S^\alpha\right)^2, \\ \mathcal{X} &:= -(S^x S^y +S^y S^z + S^z S^x),
		\end{split}
	\end{equation}
	so that $\mathcal{Z} \ket{\alpha} = \xi_\alpha \ket{\alpha}$ and $\mathcal{X} \ket{\alpha} = \ket{\beta}$, with $\xi_\beta = \omega \xi_\alpha$. Here $\xi_\alpha = 1,\omega,\omega^2$ for $\alpha=x,y,z$ respectively, and $\omega = \exp{(\ci 2\pi/3)}$. $\mathcal{X}$ acts on the operators $S^\alpha$ as $\mathcal{X} S^x \mathcal{X}^\dagger = S^y$,  $\mathcal{X} S^y \mathcal{X}^\dagger = S^z$, $\mathcal{X} S^z \mathcal{X}^\dagger = S^x$. It is convenient to state the subsequent results if we label the basis states with the eigenvalues of $\mathcal{Z}$, i.e.,~we shall write $\ket{1}, \ket{\omega}, \ket{\omega^2}$ for $\ket{x}, \ket{y}, \ket{z}$ respectively. With this, 
	\begin{equation}\label{eqn:basisV}
		\mathcal{V} = \text{span}\{ \otimes _{j} \ket{u_j}_{j \sigma} \ket{v_j}_{j \tau}, u_j, v_j \in \{ 1,\omega,\omega^2 \}\}
	\end{equation}
 \begin{figure}
	\includegraphics[width=0.99\linewidth]{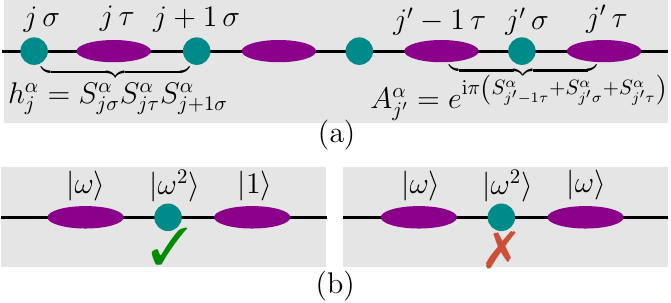}
	\caption{(a) - The one-dimensional chain on which the model is defined. Sites are shown in cyan, while the links are depicted in magenta. As per the convention used, the link $j \tau$ connects the sites $j \sigma$ and $j+1 \sigma$. The hopping terms $h^\alpha_j$ are defined on the links and the gauge transformations $A^\alpha_{j'}$ are defined on the sites. (b) - For a tensor product state in the Gauss law subspace, at any given site, the states on the site and the two links touching it must be all different. For example, the configuration shown on the left is valid, but the one on the right is not allowed as the states on the two links are the same.}
	\label{fig:model_mu} 
\end{figure}

	\paraheading{Gauss law subspace $\mathcal{V}_G$}
	We now impose the condition that the physical states of the theory must belong to  the Gauss law subspace $\mathcal{V}_G$ (with zero background charge), where $\mathcal{V}_G = \{ \ket{\psi} \in \mathcal{V}, \,\,  A^\alpha_j \ket{\psi} = \ket{\psi} \,\,  \forall j, \alpha=x,y,z \}$. The projector onto this subspace is $P_G = \prod_j \left(\frac{1+A^x_j}{2}\right) \left(\frac{1+A^y_j}{2}\right)$. We now give a description of, and construct a basis for $\mathcal{V}_G$. Since the operators $A^\alpha_j$ are all diagonal in the tensor product basis of \eqnref{eqn:basisV}, $\mathcal{V}_G$ is spanned by those product states that satisfy the Gauss law constraint. For the product state $\otimes_j\ket{u_j}_{j \sigma} \ket{v_j}_{j \tau}$ to be in $\mathcal{V}_G$, we must have (\SMRef{\secref{SM:sec:constraintsVG}}) $u_j \neq v_{j-1}, u_j \neq v_j, v_{j-1} \neq v_j$ $\forall j$. i.e.,~the states on the site $j \sigma$, and the links $j \tau$, $j-1 \,\tau$ must all be different. (see \figref{fig:model_mu}(b)).  
	A basis for $\mathcal{V}_G$ can be constructed by generating all the product states such that no two adjacent links have the same state, while the specification of states on the sites is redundant. i.e.,~$ \mathcal{V}_G  = \text{span}\{ \otimes_{j} \ket{u_j}_{j \sigma} \ket{v_j}_{j \tau}, \,\,  v_j \in\{ 1,\omega, \omega^2\}, v_j \neq v_{j-1} \,\, \text{and} \,\, u_j = v_j^\ast v_{j-1}^\ast \,\,  \forall j  \}$.
	Since the site degrees of freedom are redundant, we can effectively construct a basis for $\mathcal{V}_G$ only by specifying the states on the links. We have
	\begin{equation}\label{eqn:VGbasis}
		\mathcal{V}_G \cong \text{span}\left\{ \otimes_j \ket{v_j}_{j\tau}, v_j \in \{1,\omega,\omega^2\}, v_j \neq v_{j-1}  \forall j \right\}
	\end{equation}
	i.e.,~no two adjacent links can have the same state. This local constraint can be expressed as $\sum_{\alpha} (S^\alpha_{j \tau})^2 (S^\alpha_{j+1 \tau})^2  =1$ or $ \mathcal{Z}_{j \tau}^\dagger \mathcal{Z}_{j+1 \tau} +\mathcal{Z}_{j \tau} \mathcal{Z}_{j+1 \tau}^\dagger =-1 $ for all $j$. When $L$ is even, the dimension of $\mathcal{V}_G$  for a periodic chain of length $L$ is  $D(L) = 2^L + 2 $ (\SMRef{\secref{SM:sec:dimVG}}).  We  emphasize that even after removing the redundancy of the site degrees of freedom from the description, $\mathcal{V}_G$ \emph{does not} admit a local tensor product structure, since  the set of allowed states on a given link depends on the states on the adjacent links. The hamiltonian in \eqnref{eqn:ham}, projected to $\mathcal{V}_G$, can be effectively written in terms of the link ($\tau$) degrees of freedom (\SMRef{\secref{SM:sec:ham_u1VG}}) as
	\begin{equation}\label{eqn:hamgauss}
		H_G = -\sum_{j,\alpha} \left(t_\alpha P^\alpha_{j-1 \tau} S^\alpha_{j \tau} P^\alpha_{ j +1\tau} + K_\alpha (S^\alpha_{j \tau})^2\right),
	\end{equation}
	where $P^\alpha_{j \tau}  = 1- (S^\alpha_{j \tau})^2$. In the rest of the paper, we focus on the ground-state physics of $H_G$ subject to the local constraint $\sum_{\alpha} (S^\alpha_{j \tau})^2 (S^\alpha_{j+1 \tau})^2  =1$. In a numerical study, the  constraint may be enforced in the form of an energy penalty by adding the local term $J \sum_{j,\alpha} ((S^\alpha_{j \tau})^2 (S^\alpha_{j+1 \tau})^2 -1)$ to $H_G$, where $J>0$ and $J \gg |t_\alpha|, |K_\alpha|$. Hilbert space constraints arising from local symmetries \cite{Sen_Ramola2010} have also attracted significant interest in relation with quantum scar states and Hilbert space fragmentation \cite{Surace_Dalmonte2020,Mukherjee_Sen2021,Zhang_You2023,Mohapatra_Balram2023}.

	\paraheading{\textup{U}$(1)$ symmetry in $\mathcal{V}_G$} We now establish a key result that is crucial for the LSM theorem to be proved later: the hamiltonian $H_G$ (\eqnref{eqn:hamgauss}) in the physical subspace $\mathcal{V}_G$ enjoys a U$(1)$ symmetry. To show that $H_G$ has a U$(1)$ symmetry, we start by introducing an operator $\mu_j$ on each site $j\sigma$, that is diagonal in the basis of \eqnref{eqn:VGbasis}. Using the fact that the states $\ket{v_{j-1}}_{j-1 \tau}$ and $\ket{v_j}_{j \tau}$ on the links $j-1 \tau$ and $j \tau$ are different, we define $\pm 1$ valued $\mu_j$ as $\exp{\left(\ci \frac{2\pi}{3} \mu_j\right)} =  v_{j-1}^\ast  v_j$. $\mu_j = \pm 1$ because $v_{j-1} \neq v_j $ in the Gauss law subspace. In terms of the  $\mathcal{Z}$ (\eqnref{eqn:calZcalX}) operators,
	\begin{equation}\label{eqn:mu3def}
		\begin{split}
		& \exp{\left(\ci \frac{2\pi}{3} \mu_j \right)} = \mathcal{Z}_{j-1 \tau}^\dagger \mathcal{Z}_{j \tau}, \text{ or }\\
			&\mu_j = -\frac{\ci}{\sqrt{3}}\left(\mathcal{Z}_{j \tau} \mathcal{Z}_{j-1 \tau}^\dagger - \mathcal{Z}_{j \tau}^\dagger \mathcal{Z}_{j-1 \tau} \right).
		\end{split}
	\end{equation}
	We claim that $M:=\frac{1}{3}\sum_j \mu_j$ is a conserved quantity of  $H_G$.  It can be easily checked (\SMRef{\secref{SM:sec:ham_u1VG}}) that $M$ commutes with $H_G$. Here, we motivate the reason by considering the action of $H_G$ on a basis state. The potential term clearly commutes with $M$. Now, a key observation is that the hopping term $-P^\alpha_{j-1 \tau} S^\alpha_{j \tau} P^\alpha_{ j +1\tau}$ acts to give a nonzero ket only if $\mu_j = -\mu_{j+1}$. Under the action of this term, $\mu_j \to - \mu_j$ and  $\mu_{j+1} \to -\mu_{j+1}$, leaving $M$ invariant (see \figref{fig:u1}). Now, what are the values allowed for $M$? From  the definition \eqnref{eqn:mu3def}, we see that on a chain with periodic boundary conditions,
	\begin{equation}\label{eqn:mu3constraint}
		\exp{\left(\ci \frac{2\pi}{3} \sum_j \mu_j \right)} = \exp{\left(\ci 2\pi M \right)}  = 1.
	\end{equation}
	Thus, the conserved quantity $M$ is integer-valued and generates a U$(1)$ symmetry $U(\theta) = \exp{\left(\ci \theta M \right)}$, $\theta \in [0,2\pi)$.
	
	Let us now consider the algebra of $M$ with the spatial symmetry operators $T,R$ (\eqnref{eqn:TandR}). The translation operator $T$ acts as $T \mu_j \inv{T} = \mu_{j+1}$, so $[T,M]=0$. Whereas, we have clearly made reference to a particular direction (left to right) of the periodic chain in defining $\mu_j$. As a result, $M$ does not commute with the reflection operator $R$. In fact, we have $R \mu_j \inv{R} = -\mu_{L+2-j}$, and $RM+MR=0$. Thus, $M$ commutes with the translation operator $T$, but \emph{anticommutes} with the reflection operator $R$.
	\begin{figure}
		\includegraphics[width=0.66\linewidth]{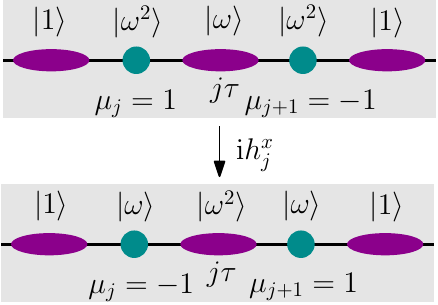}
		\caption{Conservation of $M$ is illustrated by considering the action of the hopping term $h^x_j$ on a basis state. $h^x_j$ acts to give a nonzero ket only if the links $j-1 \, \tau$ and $j+1 \, \tau$ have the state $\ket{1}$, hence only if $\mu_j=-\mu_{j+1}$. Then, under the action of this term, $\mu_j \to -\mu_j, \mu_{j+1} \to -\mu_{j+1}$, leaving $\mu_j +\mu_{j+1}$, and hence $M$ invariant.}
		\label{fig:u1}
	\end{figure}
	
	\paraheading{LSM}
	We are now ready to state and prove the main result of this work: the translation and reflection invariant hamiltonian $H_G$ of \eqnref{eqn:hamgauss} does not admit a trivial gapped ground state for any values of the parameters $t_\alpha, K_\alpha$. We prove this statement using two facts established thus far: (1)~In the Gauss law subspace, $H_G$ has a U$(1)$ symmetry, $[H_G,U(\theta)]=0$, where $U(\theta) = \exp{(\ci  \theta M)}$. (2)~The generator of the U$(1)$ symmetry, $M$, anticommutes with the reflection operator $R$.  Generally, LSM theorems arising from a global U$(1)$ symmetry   prohibit a trivial gapped ground state only when the filling-factor is not an integer \cite{Oshikawa_Affleck1997}. We emphasize that, however, in the LSM theorem obtained here, a trivial gapped ground state is prohibited for \emph{any} value of $M$ in a ground state. This is possible because of the additional property (2) listed above.
			
	The proof is simple, and proceeds as follows: Let $\ket{\psi_0}$ be a ground state of $H_G$ in the Gauss law subspace. Since $H_G$, $M$, and the translation operator $T$ all commute with each other in the Gauss law subspace, we can always choose $\ket{\psi_0}$ to also be a simultaneous eigenstate of $T$ and $M$: $H_G \ket{\psi_0} = E_0 \ket{\psi_0}$, $M \ket{\psi_0} = M_0 \ket{\psi_0}$, $T\ket{\psi_0} =e^{\ci q} \ket{\psi_0}$, $q \in [0,2\pi)$. If $M_0 \neq 0$, $\ket{\psi_0}$ is not invariant under  reflection $R$ since $M R \ket{\psi_0} = -M_0 R \ket{\psi_0}$. Thus, if $M_0 \neq 0$, $\ket{\psi_0}$ and $R \ket{\psi_0}$ are orthogonal and are eigenstates of $H_G$ with the same eigenvalue $E_0$, and our proof is complete. Hence, in the following, we can assume $M_0=0$. Now, following the standard LSM proof \cite{Oshikawa_Affleck1997}, we define a variational state $\ket{\psi}= \mathcal{O} \ket{\psi_0}$, where $\mathcal{O} = \exp{\left(\ci \frac{\pi}{L} \sum_{j =1}^L j \mu_j \right)}$. The translation operator $T$ acts on $\mathcal{O}$ as $T \mathcal{O} \inv{T} = \mathcal{O} e^{\ci \pi (\mu_1 - M/L )} = -\mathcal{O}  e^{-\ci \pi   M/L } $ as $\mu_1 = \pm 1$. Since $M_0=0$, $T \ket{\psi} = - e^{\ci q} \ket{\psi}$, so $\ket{\psi}$ is orthogonal to $\ket{\psi_0}$. Also, $\expectS{\psi}{H_G}{\psi} - E_0 \leq \frac{\text{const.}}{L}$, since $H_G$ is a local hamiltonian. Thus, the hamiltonian $H_G$ for any $t_\alpha,K_\alpha$ realizes either a spontaneously symmetry-broken, or a gapless ground state.
	
	It should be stressed that the implementation of the Gauss law is crucial for the existence of the U$(1)$ symmetry, and hence for the LSM theorem. Indeed, in the unconstrained combined Hilbert space of the matter and the gauge degrees of freedom, the hamiltonian $H=\sum_j((S^x_{j \sigma}))^2 +(S^x_{j \tau}))^2)$ admits a trivial gapped ground state $\otimes_j \ket{1}_{j \sigma} \ket{1}_{j \tau}$.
	
	Let us discuss an example to see the LSM theorem in action. Let $t_x=t_y=t_z=K_y=K_z=0$, $K_x =1$ in \eqnref{eqn:hamgauss}. The hamiltonian is $H_G = -\sum_j  (S^x_{j \tau})^2$.   For a ground state, we have $(S^x_{j \tau})^2 =1$ for all $j$. We must have $\ket{\omega}$ or $\ket{\omega^2}$ on each link. However, since adjacent links cannot have the same state, we have an exact two-fold ground state degeneracy. A ground state is $\prod_{j = 1}^{L/2} \ket{\omega}_{2j-1 \tau} \ket{\omega^2}_{2j \tau}$. The other ground state can be obtained by acting  the translation operator $T$ on the above state. We have explicitly seen an example of an SSB phase. Does the hamiltonian $H_G$ in the Gauss law subspace realize a gapless point?
	
	\paraheading{Gapless Point}
	We identify a gapless point by first mapping (\SMRef{\secref{SM:sec:mapping}}) the Gauss law Hilbert space $\mathcal{V}_G$ of dimension $D(L)=2^L+2$ to a qubit chain of length $L$ tensored with a three-dimensional Hilbert space. In this mapping, $\mu_j \equiv \mu^{(3)}_j$ will be the $j^\text{th}$ Pauli-$3$ operator of the qubit chain. Thus, due to \eqnref{eqn:mu3constraint} the qubit chain Hilbert space is constrained according to $\exp{\left(\ci \frac{2\pi}{3} \sum_j \mu^{(3)}_j\right)}=1$. The motivation behind this mapping is that a basis state of $\mathcal{V}_G$ (\eqnref{eqn:VGbasis}) can be determined by specifying the list $\{\mu_j\}$ and the state on one of the links (which is worth a three-dimensional Hilbert space). Under this \emph{exact} mapping, the hamiltonian  $H_G$ in \eqnref{eqn:hamgauss} is transformed to one that is nonlocal in the qubit operators (\cite{SM}, \eqnref{SM:eqn:hmapqubit}) for general values of the parameters $t_\alpha,K_\alpha$. However, remarkably, for $t_x=t_y=t_z$ and $K_x=K_y=K_z$, it becomes local, and can be further mapped via the Jordan-Wigner (JW) transformation to a hamiltonian that is local and quadratic in the fermion operators $c_j, c_j^\dagger$  (\cite{SM}, \eqnref{SM:eqn:Hcrit} and \eqnref{eqn:hamcrit} below). As we show in the subsequent discussion, the model is gapless at this point.  The interested reader is referred to \cite{SM}, \secref{SM:sec:mapping} for all the details concerning this mapping.
		\begin{equation}\label{eqn:hamcrit}
			H_{\text{gl}}^{(\text{f})} =  -\left( \sum_{j=1}^{L-1} \ci c_j^\dagger c_{j+1} - \ci \tilde{\mathcal{X}} c_L^\dagger c_1  e^{\ci \pi \mathcal{N}}   \right) + \text{h.c}
		\end{equation}
		The hamiltonian $H_G$ at the gapless point $t_x=t_y=t_z=1$, $K_x=K_y=K_z=0$ transforms to the hamiltonian $H_\text{gl}^{(\text{f})}$ after the mapping alluded to in the above paragraph is carried out. Here, $\tilde{\mathcal{X}} = \prod_j \mathcal{X}_{j \tau}$, with $\mathcal{X}_{j \tau}$ as given in \eqnref{eqn:calZcalX} and $[\tilde{\mathcal{X}}, c_j] = [\tilde{\mathcal{X}}, c_j^\dagger]=0$. $c_j,c_j^\dagger$ are fermion creation and annihilation operators satisfying the familiar algebra $\{c_j,c_{j'}\}=0$, $\{c_{j}, c_{j'}^\dagger\}=\delta_{j j'}$. $\mathcal{N} = \sum_j n_j$ is the total fermion number, where $n_j = c_j^\dagger c_j = (1-\mu^{(3)}_j)/2$. We have $\mathcal{N}=(L-3M)/2$ and $\exp{(\ci 2\pi M)}=1$, forcing  the total fermion number to be constrained as
		\begin{equation}\label{eqn:fermionconstraint}
			\exp{\left(\ci \frac{2\pi}{3} (L+ \mathcal{N})\right)}=1
	 	\end{equation}
	Because of this constraint, $c_j$, $c_j^\dagger$ are not valid operators in the physical Hilbert space of the system, since they change the total fermion number $\mathcal{N}$ by $\pm 1$. Let us now discuss the symmetries of the hamiltonian $H_\text{gl}^{(\text{f})}$. It  clearly conserves $\mathcal{N}$. This simply corresponds to the conservation of $M$ in the original hamiltonian $H_G$. At the gapless point, $H_\text{gl}^{(\text{f})}$  has an additional $\cyclicg{3}$ symmetry: $[H_\text{gl}^{(\text{f})}, \tilde{\mathcal{X}}] =0$, $(\tilde{\mathcal{X}})^3=1$. Indeed, this can also be inferred by inspecting the hamiltonian $H_G$ at the gapless point. $\tilde{\mathcal{X}}$ acts on the spin operators as $
		\tilde{\mathcal{X}} S^x_{j \tau} \tilde{\mathcal{X}}^\dagger = S^y_{j \tau}, \,\,	\tilde{\mathcal{X}} S^y_{j \tau} \tilde{\mathcal{X}}^\dagger = S^z_{j \tau}, \,\,	\tilde{\mathcal{X}} S^z_{j \tau} \tilde{\mathcal{X}}^\dagger = S^x_{j \tau}
	$. 
	Hence, at the gapless point, $[H_G,\tilde{\mathcal{X}}]=0$.
	We can thus label the eigenstates of $H_\text{gl}^{(\text{f})}$ with eigenvalues $1,\omega,\omega^2$ of $\tilde{\mathcal{X}}$. The spectrum of $H_\text{gl}^{(\text{f})}$ is exactly solvable because after fixing the eigenvalue of $\tilde{\mathcal{X}}$ and the parity of the total fermion number, the hamiltonian is quadratic in the fermion operators. We can do a transformation $c_j \to e^{\ci \phi_j} c_j$, so that the coefficient of $c_j^\dagger c_{j+1}$ for all $j$ becomes $e^{\ci \Phi/L}$. The flux  $\Phi$ through the periodic chain depends on the eigenvalue of $\tilde{\mathcal{X}}$ and the parity of the total fermion number $e^{\ci \pi \mathcal{N}}$ as $e^{\ci \Phi} =  e^{\ci \frac{\pi}{2} (L-1)} \times (- \tilde{\mathcal{X}} e^{\ci \frac{\pi}{2}} e^{\ci \pi \mathcal{N}} ) = - \tilde{\mathcal{X}} e^{\ci \frac{\pi L}{2}} e^{\ci \pi \mathcal{N}}$. The single particle spectrum of the $c$-fermions is given by $\epsilon(q_m) = -2 \cos{\left(q_m+\frac{\Phi}{L}\right)}$, with $q_m=\frac{2\pi}{L}m$, $m \in \{ -L/2, -L/2+1, \dots, L/2-1\}$. The many-body eigenstates have to be constructed by filling the single-particle eigenstates so that the total fermion number obeys the constraint \eqnref{eqn:fermionconstraint}.

	For the ground state, we have half-filling i.e.,~$\mathcal{N} = L/2$ (or $M=0$). Note that this is allowed as $\mathcal{N}=L/2$ satisfies the constraint \eqnref{eqn:fermionconstraint} for any even $L$. When $L$ is a multiple of $4$, we have $\tilde{\mathcal{X}}=1$ ($\Phi=\pi$) in the ground state, and the ground state is $\ket{\psi_0} =\ket{\tilde{\mathcal{X}}=1} \otimes \prod_{m =-L/4}^{L/4-1} c_{q_m}^\dagger \ket{0}$, where $c_q$ is the Fourier transform of $c_j$, and $\ket{0}$ is the vacuum of the $c$-fermions. The ground-state energy is $E_0=-2/\sin{(\pi/L)}$. When $L$ is even, but not a multiple of $4$, we have $\tilde{\mathcal{X}}= \omega$ or $\omega^2$ (two-fold degeneracy) for a ground state. This corresponds to $\Phi = 5\pi/3, \pi/3$ respectively. The ground state with $\tilde{\mathcal{X}}=\omega$ is $\ket{\psi_0} = \ket{\tilde{\mathcal{X}}=\omega} \otimes \prod_{m=-(L+2)/4}^{(L-6)/4}c_{q_m}^\dagger \ket{0}$, and $E_0= -2\cos{\left(\pi/3L\right)} \, / \sin{\left(\pi/L\right)}$. In the thermodynamic limit, $E_0/L = -2/\pi$.
	
	What are the excitations above the ground state? First, we note that states like $\ket{\psi}= c_{q}^\dagger \ket{\psi_0}$ are not allowed as they do not belong to the physical Hilbert space because of the constraint \eqnref{eqn:fermionconstraint}. However, particle-hole excitations of the $c$-fermions $c_{q'}^\dagger c_{q} \ket{\psi_0}$ describe valid excitations of the system. Here, $q$ ($q'$) is chosen such that the corresponding single-particle state is filled (empty) in the ground state. Excitations in which the number of fermions over and above the ground state is a multiple of $3$ are also allowed. $c_{q''}^\dagger c_{q'}^\dagger c_q^\dagger \ket{\psi_0}$, where the single-particle states corresponding to $q,q',q''$ are all empty in the ground state, is an example of such an excitation. Clearly, all these excitations have a vanishing gap in the thermodynamic limit $L \to \infty$ because the single-particle spectrum $\epsilon(q)$ of the $c$-fermions is gapless at half-filling. Another type of excited state can be obtained by changing the eigenvalue of $\tilde{\mathcal{X}}$. For example, when $L$ is a multiple of $4$, we can construct an excited state with $\tilde{\mathcal{X}}=\omega$. This has the effect of changing the flux $\Phi$ through the periodic chain. Hence, the energy of the state constructed this way is close to the ground state energy in the thermodynamic limit. We have thus established that the model at the point $t_x=t_y=t_z=1$, $K_x=K_y=K_z=0$ is critical, and since it admits a description in terms of free Dirac fermions, the central charge of the critical theory is $c=1$.
	
	We have used density matrix renormalization group (DMRG) to compute the bipartite entanglement entropy in the ground state of $H_G$ at the gapless point with open boundary conditions. Using the relation \cite{Calabrese_Cardy2004} between the central charge and the bipartite entanglement entropy scaling for a one dimensional critical system,  we have numerically verified that  $c \simeq 1$ (\SMRef{\secref{SM:sec:verifyc}}). The DMRG algorithm is implemented using the \texttt{ITensor} library in Julia \cite{ITensor-r0.3}.
	
	Finally, we study the behavior of correlation functions at the gapless point. $(S^\alpha_{j \tau})^2$ for $\alpha=x,y,z$ is a local gauge invariant quantity of the theory. Hence, we consider the correlation function $\expect{(S^\alpha_{j \tau})^2 (S^\beta_{j+r  \tau})^2 }$  with periodic boundary conditions in the thermodynamic limit $L \to \infty$. Making use of the  mapping developed in \cite{SM}, \secref{SM:sec:mapping} to express the spin operators in terms of $c$-fermion operators, and using the knowledge of the exact ground state, this correlation function can be expressed as the determinant  of an $r \times r$ Toeplitz matrix. Using results from the theory of Toeplitz determinants \cite{Jin_Korepin2004, Deift_Krasovsky2013}, we show (\SMRef{\secref{SM:sec:correlation}}) that the asymptotic behavior  of the correlation function is given by
	\begin{equation}\label{eqn:correlation}
				\expect{(S^\alpha_{j  \tau})^2  (S^\beta_{j+r  \tau})^2} -4/9  \sim 
					C_{\alpha \beta} \frac{\cos{(\pi r)}}{r^{2/9}},
	\end{equation}
	Note that $\expect{(S^\alpha_{j \tau})^2}=2/3$ in the ground state. In the above equation, $C_{\alpha \beta}= 2C'$, $-C'$ for $\alpha=\beta$, $\alpha \neq \beta$ respectively, and $C'$ is a positive constant.
	
	\paraheading{Some results on a spin-$1/2$ model}
	Before we conclude, we present some additional results concerning the spin-$1/2$ version of the hamiltonian in \eqnref{eqn:ham}. i.e.,~consider the hamiltonian $H_{S=1/2}= \sum_{j, \alpha}   t_\alpha \sigma^\alpha_{j} \tau^\alpha_{j} \sigma^\alpha_{j+1} $, where the sum is over $\alpha=x,y,z$ and $\sigma^\alpha_{j}$ and $\tau^\alpha_{j}$ are Pauli operators on site $j \sigma$ and link $j \tau$ respectively. This hamiltonian has local symmetries $\tilde{A}^\alpha_{j}=\tau^\alpha_{j-1} \sigma^\alpha_{j} \tau^\alpha_{j}$. Interestingly, $\tilde{A}^\alpha_{j}$ do not all commute with one another. $\tilde{A}^\alpha_j, \tilde{A}^\beta_{j'}$ anticommute with each other if $\alpha \neq \beta$ and $j-j'=0, \pm 1$, but commute otherwise.  A scenario where the local symmetries do not all commute with each other is known \cite{Pujari_Nigam2026} to give rise to an extensive degeneracy in each energy eigenvalue. Showing that the existence of these local symmetries with this algebra amounts to having a $(\cyclicg{2} \times \cyclicg{2})^L$ symmetry with a projective representation (\SMRef{\secref{SM:sec:spinhalf}}), we prove that each eigenvalue is at least $2^{L-2}$ degenerate. In this process, we develop a systematic procedure for computing degeneracies of hamiltonians with such local symmetries.

	\paraheading{Discussion}
	In this paper, we have illustrated how kinematic constraints in a gauge theory can manifest an LSM-type theorem. In the  $\cyclicg{2} \times \cyclicg{2}$ theory of matter coupled to gauge degrees of freedom we construct, imposing the Gauss law gives rise to a U$(1)$ symmetry, whose generator anticommutes with the lattice reflection operator, leading to the LSM theorem in the Gauss law subspace. The theorem forces the ground state to either spontaneously break a symmetry, or be gapless.
	
	Performing a nonlocal transformation, we have identified a gapless point, at which the model admits a description in terms of noninteracting Dirac fermions with a constraint on the fermion number. The model is a natural platform to study phase diagram topological defects \cite{Prakash_Parameswaran2023,Hsin_Thorngren2020,Manjunath_Else2601.10783} in families of SSB phases. In particular, it will be interesting to study the nature of the singularity in the phase diagram at the gapless point $t_x=t_y=t_z=1$, $K_x=K_y=K_z=0$. In the parameter space $(\delta_x,\delta_y,K_x,K_y)$, where $t_x=1+\delta_x, t_y=1+\delta_y, t_z=1-\delta_x-\delta_y$, is the gapless point an isolated critical point, or does it lie on a higher-dimensional critical manifold? It will also be interesting to explore the dynamics of the system in the constrained Hilbert space in connection with quantum scar states and Hilbert space fragmentation \cite{Mukherjee_Sen2021,Zhang_You2023,Mohapatra_Balram2023}.
	
	\paraheading{Acknowledgments}
	B.P.P acknowledges the support received through the PMRF program, Ministry of Education, India.
	\bibliography{refs}

	
	\clearpage
	\newpage
	\appendix
	
	\renewcommand{\appendixname}{}
	\renewcommand{\thesection}{{S\arabic{section}}}
	\renewcommand{\thefigure}{S\arabic{figure}}
	\renewcommand{\theequation}{\thesection.\arabic{equation}}
	
	\setcounter{page}{1}
	\setcounter{figure}{0}
	
	\begin{widetext}

		\maketitle
		
		\centerline{\bf Supplemental Material}
		\medskip
		\centerline{for}
		\medskip
		\centerline{\large{\bf \mytitle }}
		\medskip
		\centerline{by \authorOne}
		\bigskip
	\end{widetext}

	\clearpage
	\newpage

	\appendix

	\bigskip
	
\paraheading{\bf{Recap of some notation from the main text}}
On each site and link, the operators $\mathcal{Z}$ and $\mathcal{X}$ are defined as
	$
			\mathcal{Z} :=
			-\sum_{\alpha}{\xi_\alpha} \left(S^\alpha\right)^2,  \mathcal{X} := -(S^x S^y +S^y S^z + S^z S^x),
	$
	where $\xi_\alpha=1,\omega,\omega^2$ for $\alpha=x,y,z$ respectively, and $\omega=e^{\ci 2 \pi/3}$.
	
	\section{Gauss law constraints}\label{SM:sec:constraintsVG}
	Here, we derive the condition for a tensor product state to be in the Gauss law subspace. The local constraint at each site $j \sigma$ in the Gauss law subspace is $A^\alpha_j = \Sigma^\alpha_{j-1\tau}\Sigma^\alpha_{j\sigma} \Sigma^\alpha_{j\tau}=1$ for all $\alpha=x,y,z$. Here $\Sigma^\alpha = \exp{(\ci \pi S^\alpha)}= 1- 2 (S^\alpha)^2$. In terms of the projector $P^\alpha=1-(S^\alpha)^2$ onto the state $\alpha$, the above condition becomes $(2P^\alpha_{j-1 \tau} - 1) (2P^\alpha_{j \sigma} - 1) (2P^\alpha_{j \tau} - 1)=1$. Hence, $P^\alpha_{j-1\tau}+P^\alpha_{j \sigma} + P^\alpha_{j \tau} -1 = 2 P^\alpha_{j-1 \tau} P^\alpha_{j \tau} =  2 P^\alpha_{j-1 \tau} P^\alpha_{j \sigma} =  2 P^\alpha_{j \sigma} P^\alpha_{j \tau}$.	Now, using $\sum_\alpha P^\alpha =1$, we see that $ P^\alpha_{j-1 \tau} P^\alpha_{j \tau} =   P^\alpha_{j-1 \tau} P^\alpha_{j \sigma} =   P^\alpha_{j \sigma} P^\alpha_{j \tau} =0$ for all $j$ and $\alpha=x,y,z$. i.e.,~for each $j$, the states on the site $j\sigma$, and the links $j-1\, \tau$ and $j \tau$ must all be different from one another.
	
	\section{Dimension of $\mathcal{V}_G$}\label{SM:sec:dimVG}
	In this section, we calculate the dimension of the Gauss law subspace $D(L)$ for a periodic chain with $L$ sites and $L$ links. From the discussion in the main text, we see that $D(L)$ is equal to the number of ways of placing a symbol from the set $\{x,y,z\}$ on each link, so that no two adjacent links have the same symbol. Let us define $\tilde{D}(L)$ to be the total number of ways if we allow the symbols on the links $1,L$ (the first and the last link) to be equal. $\tilde{D}(L) = \sum_{\alpha,\beta} \tilde{D}_{\alpha,\beta}(L) = 3 \times 2^{L-1}$, where $\tilde{D}_{\alpha,\beta}(L)$ is the number of ways such that the first and the last link  have the symbols $\alpha$ and $\beta$ respectively. Now, we have $D(L) = \sum_{\alpha \neq \beta} \tilde{D}_{\alpha, \beta}(L) = 6 \tilde{D}_{x,y}(L) = 3 \times ( 2^{L-1} - \tilde{D}_{x,x}(L))$. We now develop a recursive relation for $\tilde{D}_{x,x}(L)$ as follows:
	$
\tilde{D}_{x,x}(L) = \tilde{D}_{y,x}(L-1) +  \tilde{D}_{z,x}(L-1) = 2 \tilde{D}_{x,y}(L-1) = 2^{L-2} - \tilde{D}_{x,x}(L-1)
	$. Using this relation again, we find that
	\begin{equation}
	\tilde{D}_{x,x}(L) - \tilde{D}_{x,x}(L-1) - 2 \tilde{D}_{x,x}(L-2) =0
	\end{equation}
	This recurrence relation has a general solution of the form $\tilde{D}_{x,x}(L) = B \times \,2^L + B' \times (-1)^L$. Using the values $\tilde{D}_{x,x}(2)=0$, $\tilde{D}_{x,x}(3)=2$, we find that $\tilde{D}_{x,x}(L) = \frac{(2^{L-1} -2 (-1)^L)}{3}$. From this, we obtain
	\begin{equation}
		D(L) = 2^L + 2 \times (-1)^L.
	\end{equation}
	
	\section{Hamiltoninan and U$(1)$ symmetry in $\mathcal{V}_G$}\label{SM:sec:ham_u1VG}
	In this section, we derive the hamiltonian for the link ($\tau$) degrees of freedom obtained after projecting the hamiltonian in \eqnref{eqn:ham}, main text to $\mathcal{V}_G$.  Consider the hopping term $h^\alpha_j= S^\alpha_{j\sigma}S^\alpha_{j \tau} S^\alpha_{j+1 \sigma}$  after projecting to $\mathcal{V}_G$. Using $S^\alpha\ket{\alpha'} = \ci \sum_{\alpha''} \epsilon^{\alpha \alpha' \alpha''} \ket{\alpha''}$, we have
	\begin{widetext}
	\begin{equation}
	 P_G h^\alpha_j P_G = (\ci)^3 \sum_{ \substack{\beta_1,\dots, \beta_L \\  \tilde{\beta}_j, \tilde{\gamma}_j, \tilde{\gamma}_{j+1} } } \epsilon^{\alpha \beta_j \tilde{\beta}_j } \epsilon^{\alpha \gamma_j \tilde{\gamma}_j} \epsilon^{\alpha \gamma_{j+1} \tilde{\gamma}_{j+1}} \ket{\gamma_1 \dots \tilde{\gamma}_j \tilde{\gamma}_{j+1} \dots \gamma_L}_\sigma \ket{\beta_1 \dots \tilde{\beta}_j \dots \beta_L}_\tau  \bra{\gamma_1 \dots \gamma_L}_\sigma \bra{\beta_1 \dots \beta_L}_\tau   
	\end{equation}
	\end{widetext}
	In the above equation, $\gamma_1, \gamma_2, \dots \gamma_L$ are all fixed according to $\xi_{\gamma_j} = \xi_{\beta_{j-1}}^\ast \xi_{\beta_j}^\ast$ due to the local constraint in $\mathcal{V}_G$. The term $ \epsilon^{\alpha \beta_j \tilde{\beta}_j } \epsilon^{\alpha \gamma_j \tilde{\gamma}_j} \epsilon^{\alpha \gamma_{j+1} \tilde{\gamma}_{j+1}} $ is nonzero only if $\alpha \neq \gamma_j, \beta_j, \gamma_{j+1}$, hence only if $\beta_{j-1}=\beta_{j+1}=\alpha$. We must also have $\gamma_j =\gamma_{j+1} \neq \alpha,\beta_j$ and $\tilde{\gamma}_j = \tilde{\gamma}_{j+1} = \beta_j$. Thus,
	\begin{widetext}
		\begin{equation}
			 P_G h^\alpha_j P_G = -\ci \sum_{ \substack{\beta_1,\dots, \beta_L \\  \tilde{\beta}_j  } } \delta_{\alpha,\beta_{j-1}} \delta_{\alpha,\beta_{j+1}} \epsilon^{\alpha \beta_j \tilde{\beta}_j }   \ket{\gamma_1 \dots \beta_j \beta_{j} \dots \gamma_L}_\sigma \ket{\beta_1 \dots \tilde{\beta}_j \dots \beta_L}_\tau  \bra{\gamma_1 \dots \gamma_L}_\sigma \bra{\beta_1 \dots \beta_L}_\tau  
		\end{equation}
	\end{widetext}
	Since the states on the sites are completely determined by those on the links, we can effectively write the term using only the link ($\tau$) degrees of freedom as $h'^{\alpha}_j= \sum_{ \{\gamma'\}, \{\gamma''\} } \bra{\{\gamma'\}}_\sigma P_G h^\alpha_j P_G \ket{\{ \gamma''\}}_\sigma$, which is equal to
	\begin{widetext}
		\begin{equation}
			h'^{\alpha}_j =  -\ci \sum_{ \substack{\beta_1,\dots, \beta_L \\  \tilde{\beta}_j  } } \delta_{\alpha,\beta_{j-1}} \delta_{\alpha,\beta_{j+1}} \epsilon^{\alpha \beta_j \tilde{\beta}_j }   \ket{\beta_1 \dots \tilde{\beta}_j \dots \beta_L}_\tau   \bra{\beta_1 \dots \beta_L}_\tau  = - P^\alpha_{j-1 \tau} S^\alpha_{j \tau} P^\alpha_{ j +1 \tau}
		\end{equation}
	\end{widetext}
	Hence, finally, the hamiltonian projected to the Gauss law subspace becomes
	\begin{equation}\label{SM:eqn:hamgauss}
		H_G = -\sum_{j,\alpha} \left(t_\alpha P^\alpha_{j-1 \tau} S^\alpha_{j \tau} P^\alpha_{ j +1\tau} + K_\alpha (S^\alpha_{j \tau})^2 \right),
	\end{equation}
where, $P^\alpha_{ j\tau} = 1-(S^\alpha_{j \tau})^2$. We now show that $M=\frac{1}{3}\sum_j \mu_j$ is a conserved quantity of the hamiltonian  $H_G$ (\eqnref{SM:eqn:hamgauss}), where $\mu_j = -\frac{\ci}{\sqrt{3}}\left(\mathcal{Z}_{j \tau} \mathcal{Z}_{j-1 \tau}^\dagger - \mathcal{Z}_{j \tau}^\dagger \mathcal{Z}_{j-1 \tau} \right)$.  First, we write  $\mu_j$ in terms of the operators $S^\alpha$. Recalling that $\mathcal{Z}_{j \tau} = -\sum_\alpha \xi_\alpha (S^\alpha_{j \tau})^2$, we get $\mu_j = - \frac{\ci}{\sqrt{3}} \sum_{\alpha, \beta} \left(\xi_\alpha \xi_\beta^\ast - \xi_\alpha^\ast \xi_\beta \right) (S^\alpha_{j \tau})^2 (S^\beta_{j-1 \tau})^2$. The potential term in \eqnref{SM:eqn:hamgauss} obviously commutes with any $\mu_j$. Hence, consider $[P^\alpha_{j-1 \tau} S^\alpha_{j \tau} P^\alpha_{ j +1\tau}, M] = \frac{1}{3} [P^\alpha_{j-1 \tau} S^\alpha_{j \tau} P^\alpha_{ j +1\tau},\mu_j +\mu_{j+1} ]=$
\begin{widetext}
	\begin{equation}
		\begin{split}
			& - \frac{\ci}{3\sqrt{3}} \sum_{\beta, \gamma} \left(\xi_\beta \xi_\gamma^\ast - \xi_\beta^\ast \xi_\gamma \right)   \left[P^\alpha_{j-1 \tau} S^\alpha_{j \tau} P^\alpha_{ j +1\tau}, (S^\beta_{j \tau})^2 (S^\gamma_{j -1 \tau})^2 +  (S^\beta_{j +1 \tau})^2 (S^\gamma_{j  \tau})^2 \right] \\
			&=-\frac{\ci}{3\sqrt{3}}  \sum_{\beta, \gamma} \left(\xi_\beta \xi_\gamma^\ast - \xi_\beta^\ast \xi_\gamma \right)   \left[P^\alpha_{j-1 \tau} S^\alpha_{j \tau} P^\alpha_{ j +1\tau}, (S^\beta_{j \tau})^2 (S^\gamma_{j -1 \tau})^2 -
			(S^\beta_{j \tau})^2 (S^\gamma_{j+1 \tau})^2  \right]=0\\
		\end{split}
	\end{equation}
\end{widetext}
	Since clearly,  $P^\alpha_{j-1 \tau} P^\alpha_{j+1 \tau} \left(  \left(S^\gamma_{j-1 \tau}\right)^2 -  \left(S^\gamma_{j+1 \tau}\right)^2 \right) = P^\alpha_{j-1 \tau} P^\alpha_{j+1 \tau} \left( P^\gamma_{j+1 \tau} - P^\gamma _{j-1 \tau}\right) =  0$.

	\section{Mapping to a qubit chain with nonlocal interactions}\label{SM:sec:mapping} In this section, we establish the mapping of the Gauss law Hilbert space $\mathcal{V}_G$ to that of a qubit chain coupled to a three-dimensional Hilbert space alluded to in the main text. We motivate the idea by the following question: Can the quantities $\{\mu_j\}$ be used to specify a basis state 
	of $\mathcal{V}_G$? We recall that  $\mu_j$ contains information about the state on the link $j$ relative to the state on the link $j-1$. We have (\eqnref{eqn:mu3def}, main text)  $\mathcal{Z}_{j \tau} = \mathcal{Z}_{j-1 \tau} \exp{\left( \ci \frac{2\pi}{3} \mu_j \right)}$. Hence, for any $j$, $\mathcal{Z}_{ j \tau} = \mathcal{Z}_{0 \tau} \exp{\left(\ci \frac{2\pi}{3} \sum_{1 \leq k \leq j} \mu_{k} \right)}$. i.e.,~The knowledge of $\mathcal{Z}_{0 \tau}$ $(\equiv \mathcal{Z}_{L \tau})$ and $\{\mu_k\}$ is sufficient to determine all other $\mathcal{Z}_{j \tau}$ for $j=1,2 \dots L-1$, and hence the basis state of $\mathcal{V}_G$. For a periodic chain, the $\{\mu_j\}$  must satisfy the constraint
	\begin{equation}\label{SM:eqn:constraint}
		\exp{\left(\ci \frac{2\pi}{3} \sum_j \mu_j\right)} = e^{\ci 2\pi M} =1.
	\end{equation} Hence, we have the mapping
	\begin{equation}
		\begin{split}
			&\mathcal{V}_G \cong \text{span}\left\{ \ket{v_0; \mu_1', \dots ,\mu_L'},\,\, v_0 \in \{1,\omega,\omega^2 \},\,\, \mu_j' = \pm 1, \right. \\
			& \qquad \qquad \qquad \qquad \qquad \qquad \left.\exp{\left(\ci \frac{2\pi}{3} \sum_j \mu_j' \right)} =1 \right\}\\
			&\mu_j \ket{v_0; \mu_1', \dots ,\mu_L'} = \mu_j' \, \ket{v_0; \mu_1', \dots ,\mu_L'} \\
			&\mathcal{Z}_{0 \tau} \ket{v_0; \mu_1',  \dots ,\mu_L'} = v_0 \, \ket{v_0; \mu_1', \dots, \mu_L'}
		\end{split}
	\end{equation}
	 We have mapped $\mathcal{V}_G$ to the Hilbert space of a chain of qubit degrees of freedom subject to the above constraint,  tensored with a degree of freedom worth a three-dimensional Hilbert space. $\mu_j \equiv \mu^{(3)}_j$ will be the Pauli-$3$ operator associated with the $j^\text{th}$ qubit. We note that this mapping is consistent with the dimension of $\mathcal{V}_G$ calculated in \secref{SM:sec:dimVG}, since the number of solutions to the above constraint with $\mu_j' = \pm 1$ is equal to $(2^L +  2  \times (-1)^L )/3$, giving us $D(L) = 2^L +2 \times (-1)^L$ as before.
	 
	 Let us now discuss the operator content on this side of the mapping. The Pauli-$3$ operators $\mu^{(3)}_j$ and $\mathcal{Z}_{0 \tau}$ are valid operators of the theory. We define the Pauli-$1$ and Pauli-$2$ operators $\mu^{(1)}, \mu^{(2)}$ via their action on the eigenstates of $\mu^{(3)}$, $\{ \ket{\mu}, \mu=\pm 1 \}$, as $\mu^{(1)} \ket{\mu} = \ket{-\mu}$, $\mu^{(2)} \ket{\mu} = (\ci)^\mu \ket{-\mu}$. Importantly, the  operators $\mu^{(1)}_j$ and $\mu^{(2)}_j$ are not allowed in the Gauss law Hilbert space $\mathcal{V}_G$ since they do not respect the constraint \eqnref{SM:eqn:constraint}: these operators change $M=\frac{1}{3}\sum_j \mu_j$ by $\pm 2/3$. However, as we will see below, combinations like $\mu^{(1)}_j \mu^{(2)}_{j+1} - \mu^{(2)}_j \mu^{(1)}_{j+1}$ are allowed. Now, what is an operator that acts as a ``cyclic raising operator" on $\mathcal{Z}_{0 \tau}$, but leaves all $\mu^{(3)}_j$ invariant? Clearly, such an operator must act as a raising operator on all $\mathcal{Z}_{j \tau}$. It is $\tilde{\mathcal{X}} = \prod_j (\mathcal{X}_{j \tau})$. We have $ \mathcal{Z}_{ 0 \tau} \tilde{\mathcal{X}} = \omega \tilde{\mathcal{X}} \mathcal{Z}_{ 0 \tau}$ and $[\tilde{\mathcal{X}}, \mu^{(s)}_j ] = 0$ for $s=1,2,3$. Now, we are ready to see how the hamiltonian of \eqnref{SM:eqn:hamgauss} transforms under this mapping. 
	 
	 First, consider the potential term $(S^\alpha_{j \tau})^2 = 1 - P^\alpha_{ j\tau}$, where $P^\alpha_{ j\tau} = (1+\xi_\alpha^\ast \mathcal{Z}_{j \tau} + \xi_\alpha \mathcal{Z}_{j \tau}^\dagger)/3$ is the projector on to the state $\alpha$ on the link $j \tau$. Hence we have
	 \begin{equation}\label{SM:eqn:projector}
	 	\begin{split}
	 	 P^\alpha_{j \tau} &= 1- (S^\alpha_{j \tau})^2 \\ &=\frac{1}{3} + \frac{1}{3} \left(\xi_\alpha^\ast \mathcal{Z}_{0 \tau} \exp{\left(\ci \frac{2\pi}{3} \sum_{1 \leq k \leq j} \mu^{(3)}_{k}\right)} + \text{h.c}\right)
		\end{split}
	 \end{equation}
	Now, consider the action of the hopping term ${h'}^\alpha_j = - P^\alpha_{j-1 \tau} S^\alpha_{j \tau} P^\alpha_{ j +1 \tau} $ for $1 \leq j < L$ on the state $\ket{\psi}= \ket{v_0;  \mu_1' \dots \mu_L'}$. ${h'}^\alpha_j \ket{\psi} \neq 0$ only if $P^\alpha_{j-1 \tau} \ket{\psi} = P^\alpha_{j+1 \tau} \ket{\psi} = \ket{\psi}$. Or, equivalently, only if $P^\alpha_{j-1 \tau} \ket{\psi}= \ket{\psi}$ and $\mu_j' +\mu_{j+1}'=0$.  We have
${h'}^\alpha_j \ket{v_0 ; \mu_1' \dots \mu_j' \mu_{j+1}' \dots \mu_L'} = P^\alpha_{j-1 \tau} (-\ci \mu_j') \delta_{\mu_j', -\mu_{j+1}'} \ket{v_0 ; \mu_1' \dots \mu_{j+1}' \mu_{j}' \dots \mu_L'} $ for $1 \leq j < L$. Hence,
\begin{equation}
	{h'}^\alpha_j = \frac{1}{2} P^\alpha_{j-1 \tau}    \left(\mu^{(1)}_j \mu^{(2)}_{j+1} - \mu^{(2)}_j \mu^{(1)}_{j+1}\right), \,\,\, 1 \leq j < L
\end{equation}
	where $P^\alpha_{j \tau}$ is defined in \eqnref{SM:eqn:projector}. The case ${h'}^\alpha_L$ must be  treated separately since it can change the eigenvalue of $\mathcal{Z}_{ 0 \tau} \equiv \mathcal{Z}_{ L \tau}$. We have ${h'}^\alpha_L \ket{v_0 ; \mu_1' \mu_2' \dots \mu_{L-1}' \mu_L'} = P^\alpha_{L-1}
	(-\ci \mu_L') \delta_{\mu_L', -\mu_1'} \ket{ \omega^{-\mu_1'} v_0; \mu_L' \mu_2' \dots \mu_{L-1}' \mu_1'}$, hence,
	\begin{equation}
		h^\alpha_L = \frac{1}{2} P^\alpha_{L-1} \left( \mu^{(1)}_L \mu^{(2)}_{1} - \mu^{(2)}_L \mu^{(1)}_{1} \right) \tilde{\mathcal{X}}^{-\mu_1}.
	\end{equation}
	Hence, under the mapping to qubit operators, the hamiltonian \eqnref{SM:eqn:hamgauss} (up to the addition of a constant term) transforms to
	\begin{widetext}
	\begin{equation}\label{SM:eqn:hmapqubit}
		H_G^{(\text{q})} = \frac{1}{2}  \sum_{j=1}^{L-1}  \sum_{\alpha} t_\alpha P^\alpha_{j-1 \tau}   \left(\mu^{(1)}_j \mu^{(2)}_{j+1} - \mu^{(2)}_j \mu^{(1)}_{j+1}\right) + \frac{1}{2}  \sum_{\alpha} t_\alpha P^\alpha_{L-1 \tau}  \left( \mu^{(1)}_L \mu^{(2)}_{1} - \mu^{(2)}_L \mu^{(1)}_{1} \right) \tilde{\mathcal{X}}^{-\mu_1}  + \sum_{j, \alpha} K_\alpha P^\alpha_{j \tau}
	\end{equation}
	\end{widetext}
	where $P^\alpha_{j \tau}$ consists of an operator string involving the operators $\{\mu^{(3)}_j\}$ as given in \eqnref{SM:eqn:projector}. We note that for general values of the parameters $t_\alpha,K_\alpha$, the hamiltonian \eqnref{SM:eqn:hmapqubit} is not local in terms of the qubit operators $\{\mu^{(s)}\}, s=1,2,3$. Let us do a Jordan-Wigner (JW) transformation on the  qubit operators $\mu^{(s)}_j, s=1,2,3$. The JW transformation is
	\begin{equation}
		\begin{split}
				\mu^{(1)}_j &= (c_j + c_j^\dagger)  \prod_{1 \leq k < j} \mu^{(3)}_k \\
				\mu^{(2)}_j &= -\ci (c_j -c_j^\dagger)  \prod_{1 \leq k < j} \mu^{(3)}_k
		\end{split}
	\end{equation}
	Then $\mu^{(3)}_j = 1-2n_j$, where $n_j = c_j^\dagger c_j$ is the fermion number. Let $\mathcal{N} = \sum_j n_j = (L-3M)/2$ be the total fermion number operator. The constraint \eqnref{SM:eqn:constraint} expressed in terms of the fermion operators then becomes
	\begin{equation}\label{SM:eqn:fermionconstraint}
		\exp{\left(\ci \frac{2\pi}{3}( L + \mathcal{N}) \right)}  =1
	\end{equation}
	The fermion creation and annihilation operators $c_j,c_j^\dagger$ are not valid operators in the theory because of this constraint. The hamiltonian \eqnref{SM:eqn:hmapqubit} after JW transforms to
	\begin{widetext}
	\begin{equation}\label{SM:eqn:fermionmapping}
		H_G^{(\text{f})} = \sum_{j=1}^{L-1} \sum_{\alpha} t_\alpha  P^\alpha_{j -1 \tau} \left(-\ci c_j^\dagger c_{j+1} + \ci c_{j+1}^\dagger c_j\right)  + \sum_\alpha t_\alpha P^\alpha_{L-1 \tau} \left(\ci \tilde{\mathcal{X}} c_L^\dagger c_1 - \ci \tilde{\mathcal{X}}^\dagger c_1^\dagger c_L\right) \exp{\left(\ci \pi \mathcal{N}\right)} + \sum_{j,\alpha} K_{\alpha} P^\alpha_{j \tau}
	\end{equation}
	\end{widetext}
	The appearance of the factor $\exp{(\ci \pi \mathcal{N})}$ in the $L^\text{th}$ term when performing JW transformation with periodic boundary conditions is well known. The projector $P^\alpha_{j \tau}$ in terms of the fermion operators is
	\begin{equation}\label{SM:eqn:fermionprojector}
	 P^\alpha_{j \tau} = \frac{1}{3} + \frac{1}{3} \left(\xi_\alpha^\ast \mathcal{Z}_{0 \tau} e^{\ci \frac{2\pi j}{3}} \exp{\left(\ci \frac{2\pi}{3} \sum_{1 \leq k \leq j} n_k\right)} + \text{h.c}\right)
	\end{equation}
	This exact mapping to the fermionic hamiltonian offers us another way to see the existence of the conserved quantity $M = (L - 2 \mathcal{N})/3$  in the Gauss law subspace, since the above hamiltonian in \eqnref{SM:eqn:fermionmapping} clearly conserves the total fermion number $\mathcal{N}$.
	
	When $t_x=t_y=t_z(=1)$ and $K_x=K_y=K_z(=0)$, because of the identity $\sum_\alpha P^\alpha_{j \tau}=1$, $H_G^{(\text{q})}$ is local in the qbit operators. At this point, $H_G^{(\text{f})}$ is also local and quadratic in the fermion operators. They are given by
	\begin{widetext}
	\begin{equation}\label{SM:eqn:Hcrit}
		\begin{split}
		H_{\text{gl}}^{(\text{q})}&= \frac{1}{2} \sum_{j=1}^{L-1} \left(\mu^{(1)}_j \mu^{(2)}_{j+1} - \mu^{(2)}_j \mu^{(1)}_{j+1}\right) + \frac{1}{2}  \left( \mu^{(1)}_L \mu^{(2)}_{1} - \mu^{(2)}_L \mu^{(1)}_{1} \right) \tilde{\mathcal{X}}^{-\mu_1} \\
		H_{\text{gl}}^{(\text{f})}&= \sum_{j=1}^{L-1} \left(-\ci c_j^\dagger c_{j+1} + \ci c_{j+1}^\dagger c_j\right)  +  \left(\ci \tilde{\mathcal{X}} c_L^\dagger c_1 - \ci \tilde{\mathcal{X}}^\dagger c_1^\dagger c_L\right) e^{\ci \pi \mathcal{N}}
		\end{split}
	\end{equation}
	\end{widetext}

	\section{Numerical verfication of the central charge}\label{SM:sec:verifyc}
\begin{figure}
		\includegraphics[width=0.99\linewidth]{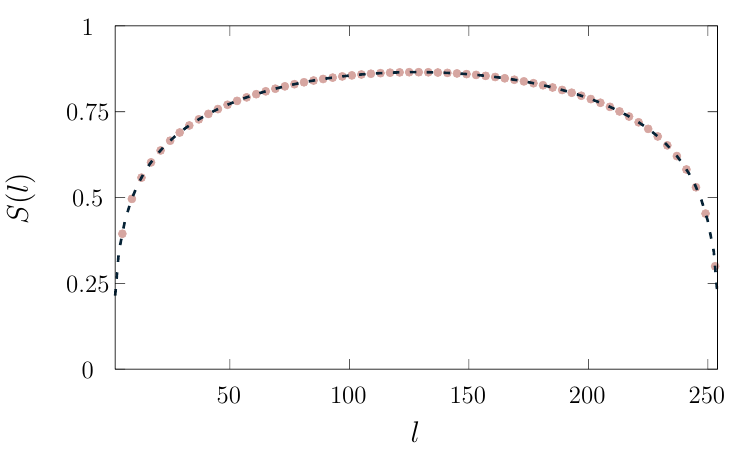}
		\caption{Plot of entanglement entropy $S(l)$ vs the subsystem size $l$ for a system of size $L=256$ (data for every four points is shown). The dashed line shows the fit with the entanglement scaling formula, which gives $c \simeq 1$}
		\label{SM:fig:EE}
	\end{figure}

	We have numerically computed the central charge at the gapless point $t_x=t_y=t_z=1$, $K_x=K_y=K_z=0$ of $H_G$ (\ref{SM:eqn:hamgauss}) by studying the bipartite entanglement entropy in the ground state with open boundary conditions using density matrix renormalization group (DMRG). The local constraint is imposed as an energy penalty by adding the term $J \sum_{\alpha} ((S^\alpha_{j \tau})^2 (S^\alpha_{j+1 \tau})^2 -1 )$, where $J>0, J \gg 1$, to the hamiltonian $H_G$. It is known \cite{Calabrese_Cardy2004} that for a one-dimensional critical system with open boundary conditions, the von Neumann entanglement entropy for the partition into subsystems of sizes $l$ and $L-l$, both of which contain an edge, is $S_{\text{vN}}(l) = \frac{c}{6}  \ln{((2L/\pi) \sin{(\pi l/L)})} +c'$. We perform DMRG on a system of size $L=256$ and extract $c \simeq 1$ (see \figref{SM:fig:EE}).

	\section{Correlation functions at the gapless point}\label{SM:sec:correlation}
	As discussed in the main text, to diagonalize the hamiltonian at the gapless point, we perform a transformation $c_j \to e^{\ci \phi_j} c_j$ so that the coefficient of $c_j^\dagger c_{j+1}$ is equal to $e^{\ci \Phi/L}$ for all $j$. The spin correlation functions we calculate below involve only the fermion particle numbers $n_j$, which are anyway invariant under this transformation. Hence, we can work with the correlation functions of the transformed $c$ operators.
	\subsection{Correlation function as a Toeplitz determinant}
In the ground state, consider the correlation function $\expect{c_j^\dagger c_k} = \frac{1}{L} \sum_{q} e^{-\ci q ( j - k)} \expect{n_q}$. Clearly, $\expect{c_j^\dagger c_j}=1/2$. For $L$ a multiple of $4$, $\expect{c_j^\dagger c_k} = \frac{1}{L} \sum_{m=-L/4}^{L/4-1} e^{-\ci \frac{2\pi m}{L}(j-k)}.$ For $j\neq k$,
	\begin{equation}
		\begin{split}
	&\expect{c_j^\dagger c_k} = \frac{\sin{\left(\frac{\pi}{2}(j-k)\right)}}{L \sin{\left(\frac{\pi }{L} (j-k)\right)}} e^{\ci \pi (j-k)/L} \\
	& \expect{c_j^\dagger c_k} - \expect{c_k^\dagger c_j}  = \frac{2\ci}{L} \sin{\left(\frac{\pi}{2}(j-k)\right)}
	\end{split}
	\end{equation}
	If $L$ is even, but not a multiple of $4$ (for one of the ground states), $\expect{c_j^\dagger c_k} = \frac{1}{L} \sum_{m=-(L+2)/4}^{(L-6)/4} e^{-\ci \frac{2\pi m}{L}(j-k)}$, and for $j\neq k$
		\begin{equation}
		\begin{split}
			\expect{c_j^\dagger c_k} &= \frac{\sin{\left(\frac{\pi}{2}(j-k)\right)}}{L \sin{\left(\frac{\pi }{L} (j-k)\right)}} e^{ 2\ci \pi (j-k)/L} \\
			 \expect{c_j^\dagger c_k} - \expect{c_k^\dagger c_j}  &= \frac{4\ci}{L} \sin{\left(\frac{\pi}{2}(j-k)\right)} \cos{\left(\frac{\pi}{L} (j-k)\right)}
		\end{split}
	\end{equation}
	In any case, in the thermodynamic limit $L \gg |j-k|$, (we first take the limit $L \to \infty$), the correlation function
	\begin{equation}\label{SM:eqn:expect_cdagc}
		\expect{c_j^\dagger c_k} =  \left\{ \begin{array}{c c}
			1/2 & \text{if } j=k \\
			\frac{\sin{\left(\frac{\pi}{2}(j-k)\right)}}{\pi (j-k)} & \text{if } j \neq k
		\end{array}  \right.
	\end{equation}
	Let us express the spin correlation functions in terms of the fermion operators. We have $(S^\alpha_{j \tau})^2- 2/3 = -(\xi_\alpha^\ast \mathcal{Z}_{j \tau} + \xi_\alpha \mathcal{Z}_{j \tau}^\dagger )/3$, where $\mathcal{Z}_{j \tau} = \mathcal{Z}_{0 \tau} e^{\ci \frac{2\pi}{3} j} \exp{\left(\ci \frac{2\pi}{3}  \sum_{1 \leq k \leq j} n_k \right) }$ (\eqnref{SM:eqn:fermionprojector}). At the gapless point, the ground state is an eigenstate of $\tilde{\mathcal{X}}$, hence $\expect{\mathcal{Z}_{ 0 \tau}}=\expect{\mathcal{Z}_{j \tau}}=0$ and $\expect{(S^\alpha_{j \tau})^2}=2/3$ in the ground state. Now, consider
	\begin{equation}\label{SM:eqn:correlationfn}
		\begin{split}
		&\expect{ (S^\alpha_{j \tau})^2    (S^\beta_{j+r \tau})^2 }-4/9= \\
	& \quad \quad \quad \quad \quad \quad \quad \frac{1}{9} \left( \xi_\alpha^\ast \xi_\beta \expect{\mathcal{Z}_{j \tau}\mathcal{Z}_{j +r\tau}^\dagger }  + \text{c.c} \right).
		\end{split}
	\end{equation}
	We have
	\begin{align}
		\expect{\mathcal{Z}_{j \tau}\mathcal{Z}_{j +r\tau}^\dagger} = e^{\ci \frac{2\pi}{3} r} \expect{ \exp{\left(\ci \frac{2\pi}{3} \sum_{j+1 \leq k \leq j+r} n_k \right)}}
	\end{align}
	We can write $\exp{\left(\ci \frac{2\pi}{3} n_k \right)} = (c_k + \omega c_k^\dagger) (c_k +c_k^\dagger)$. Defining $a_k = c_k + \omega c_k^\dagger$, $b_k= c_k +c_k^\dagger$,
	we have 
	\begin{align}
		\expect{a_k a_l}&= \omega \expect{b_k b_l} =\omega \delta_{kl} +\omega\expect{c_k^\dagger c_l } - \omega\expect{ c_l^\dagger c_k }  \\
		\expect{a_k b_l} &= \delta_{kl} +\omega  \expect{c_k^\dagger c_l } - \expect{c_l^\dagger c_k} 
	\end{align}
Using \eqnref{SM:eqn:expect_cdagc}, we have in the thermodynamic limit $L \to \infty$ ($L \gg |k-l|$),
	\begin{equation}\label{SM:eqn:expect_ab}
		\begin{split}
			\expect{a_k a_l}&= \omega \expect{b_k b_l}= \omega \delta_{kl} \\
		\expect{a_k b_l} &=  \left\{ \begin{array}{c c}
				\frac{\omega+1}{2} & \text{if } k=l \\
				 \frac{(\omega-1)}{\pi (k-l)}  \sin{\frac{\pi}{2} (k-l) } & \text{if } k \neq l
			\end{array}  \right.
		\end{split}
	\end{equation}
	Now, $\expect{\mathcal{Z}_{j \tau}\mathcal{Z}_{j +r\tau}^\dagger} = e^{\ci \frac{2\pi}{3} r} \expect{\prod_{k=j+1}^{ j+r} (a_k b_k)}$. Using Wick's theorem and \eqnref{SM:eqn:expect_ab}, we can write in the thermodynamic limit
$\expect{\mathcal{Z}_{j \tau}\mathcal{Z}_{j +r \tau}^\dagger} = e^{\ci \frac{2\pi r}{3}} \text{det}(F')$, where $F'$ is the $r \times r$  Toeplitz matrix with the entries $F'_{k,l} = F'_{k-l}=\expect{a_k b_l}$ for $0 \leq k,l \leq r-1$. Factoring out $e^{\ci \frac{2\pi}{3}} e^{\ci \frac{\pi}{2} k}$ from every row $k$ and $e^{-\ci \frac{\pi}{2} l}$ from every column $l$, $\expect{\mathcal{Z}_{j \tau}\mathcal{Z}_{j +r\tau}^\dagger} = \cos{(\pi r)} \,  \text{det}(F)$, where $F$ is a real Toeplitz matrix, $F_{k,l} = F_{k-l}$ with
\begin{align}\label{SM:eqn:Fmatrix}
	F_{k-l} = \left\{ \begin{array}{c c}
		1/2 & \text{if } k=l \\
		 \frac{\sqrt{3}}{\pi (k-l) }  \sin^2{\left(\frac{\pi}{2} (k-l) \right) }  & \text{if } k \neq l
	\end{array}  \right.
\end{align}
$\text{det}(F)$ is real since $F$ is real.

\subsection{Asymptotic limit}
To understand the asymptotic behavior of the correlation function when the separation between the lattice points $r \to \infty$ (note we have already taken the limit $L \to \infty$), we need to know the asymptotic behavior of $\text{det}(F)$ as the dimension of the matrix $r \to \infty$. We now state the Fisher-Hartwig conjecture (see Ref.~\cite{Deift_Krasovsky2013} for a review), which can be used to determine the asymptotic behavior of Toeplitz determinants.

Let $f:S^1 \to \mathbb{C}$, and let $\{F_p\}$ be the Fourier coefficients of $f$:
\begin{equation}
	\begin{split}
	f(\theta) &= \sum_{p=-\infty}^{\infty} F_p e^{\ci p \theta}\\
	F_p &= \frac{1}{2\pi} \int_0^{2\pi} \text{d}\theta f(\theta) e^{-\ci p \theta}
	\end{split}
\end{equation}
Let $f(\theta)$, with $z=e^{\ci \theta}$, be of the form
\begin{equation}\label{SM:eqn:Fisher-Hartwig}
	f(\theta) = e^{W(z)} z^{\sum_{s=1}^Q \eta_s} \prod_{s=1}^Q |z-z_s |^{2 \nu_s} w_{z_s, \eta_s}(\theta) z_s^{-\eta_s},
\end{equation}
where
\begin{equation}
	w_{z_s, \eta_s}(\theta) = \left\{  \begin{array}{cc}
		e^{\ci \pi \eta_s}, 0 \leq \theta < \theta_s \\
		e^{-\ci \pi \eta_s}, \theta_s \leq \theta < 2\pi
	\end{array}\right.,
\end{equation}
and $W: S^1 \to \mathbb{C}$ is a smooth function. Also, $z_s=e^{\ci \theta_s}$, $0<\theta_s <2\pi$, $\eta_s \in \mathbb{C}$, $\text{Re}(\nu_s) > -1/2$ for $s=1,\dots Q$. Then, according to the Fisher-Hartwig conjecture, the determinant $\text{det}(F^{(r)})$ of the $r \times r$ Toeplitz matrix $F^{(r)}_{k,l}=F_{k-l}, 0 \leq k,l \leq r-1$ has the asymptotic form
\begin{equation}\label{SM:eqn:FHconjecture}
	\text{det}(F^{(r)}) \sim C \times e^{r W_0} \times r^{\sum_s (\nu_s^2 - \eta_s^2)},
\end{equation}
where $C$ is a constant, $W_0= \frac{1}{2\pi} \int_0^{2\pi} \text{d}\theta \, W(e^{\ci \theta})$.  This conjecture has been proven for $|\text{Re}({\eta_s})| < 1/2, |\text{Re}({\nu_s})| < 1/2$, which includes the case in which we use this below. Now, for the Toeplitz matrix \eqnref{SM:eqn:Fmatrix} we are concerned with,
\begin{equation}
		\begin{split}
	f(\theta) &= 
	 \frac{1}{2} + \frac{\sqrt{3}}{\pi} \sum_{p=1}^{\infty} \frac{1}{p} \sin^2{\left(\frac{\pi p}{2}\right)} \left(e^{\ci p \theta } - e^{- \ci p \theta}\right)\\
	 &= \frac{1}{2} + \frac{\sqrt{3}}{\pi} \left(\left(\sum_{p=1}^{\infty} \frac{z^{2p-1}}{2p-1}\right) -
	\left(\sum_{p=1}^{\infty} \frac{z^{1-2p}}{2p-1}\right) \right) \\
	&= \frac{1}{2} + \frac{\sqrt{3}}{2\pi}
	(\log{(1+z)} - \log{(1-z)} \\ & \quad \quad \quad \quad \quad- \log{(1+1/z)} + \log{(1-1/z)})
	\end{split}
\end{equation}
Which gives
\begin{align}
	f(\theta) = \left\{ \begin{array}{c c}
		e^{\ci \frac{\pi}{3}} & \,\, 0 < \theta < \pi \\
		e^{- \ci \frac{\pi}{3}}  & \,\, \pi < \theta < 2\pi \\
	\end{array}  \right.
\end{align}
$f$ can be cast in the Fisher-Hartwig form \eqnref{SM:eqn:Fisher-Hartwig} with $W(e^{\ci \theta})=0$, $Q=2$, $\eta_1=1/3,\eta_2 = - 1/3$, $\nu_1 = \nu_2 =0$, $\theta_1=\pi, \theta_2=0$:
\begin{align}
	f(\theta) = w_{-1,1/3}(\theta) w_{1,-1/3}(\theta)
\end{align}
 Using \eqnref{SM:eqn:FHconjecture}, we conclude that asymptotically, for large $r$, $\text{det}(F) \sim C \frac{1}{r^{2/9}}$. Hence, $\expect{\mathcal{Z}_{j \tau}\mathcal{Z}_{j +r\tau}^\dagger} \sim C \frac{\cos{(\pi r)}}{r^{2/9}}$. Using the formula given for the constant $C$ in Ref.~\cite{Deift_Krasovsky2013},  $C=\left(2^{-1/9}G(4/3) G(2/3)\right)^2$, where $G(z)$ is the Barnes-G function, is a positive constant. Finally, from \eqnref{SM:eqn:correlationfn}, we have $\expect{ (S^\alpha_{j \tau})^2   (S^\beta_{j+r \tau})^2 } -4/9 \sim C_{\alpha \beta} \frac{\cos{(\pi r)}}{r^{2/9}}$, where $C_{\alpha \beta}= 2C/9$,  $-C/9$ for $\alpha=\beta$, $\alpha \neq \beta$ respectively.
 
 \section{Spin-$1/2$ model}\label{SM:sec:spinhalf}
 The spin-$1/2$ model we consider is
 \begin{equation}
	H_{S=1/2}= \sum_{j, \alpha}   t_\alpha \sigma^\alpha_{j} \tau^\alpha_{j} \sigma^\alpha_{j+1}
	\end{equation}
 with periodic boundary conditions. This hamiltonian has local symmetries $\tilde{A}^\alpha_{j}=\tau^\alpha_{j-1} \sigma^\alpha_{j} \tau^\alpha_{j}$. These operators do not all commute with each other. If $\alpha \neq \beta$ and $j'-j =0,\pm 1$, $\tilde{A}^\alpha_j, \tilde{A}^\beta_{j'}$ anticommute. Otherwise, they commute. We show this implies that the symmetry group $\biggroup = (\cyclicg{2} \times \cyclicg{2})^L$ is realized as a projective representation on the Hilbert space.
 
 Let $\cyclicg{2} \times \cyclicg{2}=\{e,a,b,c=ab\}$. 
 Consider the projective representation of $\cyclicg{2} \times \cyclicg{2}$  on the Hilbert space of each site and link with the multiplier system $\omega_0$: $u_{j \sigma}(e) = 1 , u_{j \sigma}(a) = \ci \sigma^x_j, u_{j \sigma}(b)=\ci \sigma^y_j, u_{j \sigma}(c) = \ci \sigma^z_j$,   $u_{j \tau}(e) = 1 , u_{j \tau}(a) = \ci \tau^x_j, u_{j \tau}(b)=\ci \tau^y_j, u_{j \tau}(c) = \ci \tau^z_j$. The  symmetry group associated with the set of local symmetries discussed above is $\biggroup \cong (\cyclicg{2} \times \cyclicg{2})^L$. Let $\elembigg{g}=\left(g_1, g_2, \dots g_L\right) \in \biggroup$, where each $g_j \in \cyclicg{2} \times \cyclicg{2}$. The action of the symmetry group is
 \begin{equation}\label{SM:eqn:projrep}
 	U(\elembigg{g}) = \prod_j \left(u_{ j \sigma}(g_j) u_{ j \tau} (g_j g_{j+1}) \right)
 \end{equation}
 and the multiplier system associated with this representation is
 \begin{equation}
 	\Omega(\elembigg{g}, \elembigg{h}) = \prod_j\left(\omega_0\left(g_j, h_j\right) \omega_0\left(g_j g_{j+1}, h_j h_{j+1} \right)\right).
 \end{equation}
 The object of interest for us is the subgroup ${\cal M} < \biggroup$ defined as
\begin{align}
	{\cal M} = \{\elembigg{g} \in \biggroup \,\, | \,\, [U(\elembigg{g}), U(\elembigg{h}) ] =0 \,\, \forall \elembigg{h} \in \biggroup \}
\end{align}
For a certain class of projective representations which includes the present case, we show below that the dimension of any irreducible representation of the projective representation is $d= \sqrt{|\biggroup|/|\mathcal{M}|}$. In the present case, $(g_1, g_2, \dots g_L) \in {\cal M} \implies g_{j-1}g_j g_{j+1} = 1$ for all $j$. Hence, if $L$ is not a multiple of $3$, then ${\cal M} = 1$ is the trivial group, and the degeneracy of each eigen space is (at least) $2^L$. If $L$ is a multiple of $3$, then ${\cal M} = \{ \left(g_1, g_2, g_1g_2, g_1, g_2, g_1g_2, \dots, g_1, g_2, g_1g_2\right) | g_1, g_2 \in \cyclicg{2} \times \cyclicg{2} \}$, $|{\cal M}| = 4^2$, and the dimension of any eigen space is at least $2^{L-2}$.

\subsection{Dimension of irrep}
	Let $\biggroup \cong \cyclicg{2}^N$ for some $N$. Then the Schur multiplier $\schurmult{\biggroup} \cong \cyclicg{2}^{N(N-1)/2}$. Let $U$ be a projective representation of $\biggroup$ on a Hilbert space with $\Omega$ as the associated multiplier system.  Also, let $\Omega(\elembigg{g}, \elembigg{h})^2 = 1$ $\forall \elembigg{g}, \elembigg{h} \in \biggroup$. We want to find the irreducible representations contained in the direct sum decomposition of $U$.
	
	Now, the projective representation $U$ can be lifted to a linear representation $\rho$ of a central extension $X$ of $\biggroup$ by $V \cong \cyclicg{2} = \{ E,F\}$. i.e., $X/V = \biggroup$. As a set, $X= \{ (\elembigg{g},a) | \elembigg{g} \in \biggroup, a \in \cyclicg{2} \}$. The group multiplication in $X$ is
\begin{align}
	(\elembigg{g}, a_1) (\elembigg{h}, a_2) &= (\elembigg{g} \elembigg{h}, a_1 a_2 \alpha(\elembigg{g}, \elembigg{h})) \\
	\alpha(\elembigg{g},\elembigg{h}) &= \left\{  \begin{array}{l l}
		E &  \text{if } \Omega(\elembigg{g}, \elembigg{h}) = 1 \\
		F &  \text{if } \Omega(\elembigg{g}, \elembigg{h}) = -1
	\end{array}\right.
\end{align}
$\alpha$ is an element of the second cohomology group $\secondcohomology{\biggroup}{{\cyclicg{2}}}$. The linear representation $\rho$ of $X$ is $\rho((\elembigg{g},a)) = \lambda(a) U(\elembigg{g})$, where $\lambda$ is the nontrivial linear representation of $\cyclicg{2}$; $\lambda : E\mapsto 1, F \mapsto -1$.

We want to find the dimensions of linear irreducible representations of $X$ that descend to a projective representation of $\biggroup$ with the same multiplier system $\Omega$. Let $\rho_0$ be a linear irrep of $X$. Any irreducible linear representation when restricted to the group center acts as a scalar representation (due to Schur's lemma). In fact, the same is true for any subgroup of the center. If a linear irrep $\rho_0$ of $X$ has to descend to a projective irrep of $\biggroup$ with the multiplier system $\Omega$, then it should act as the nontrivial linear representation $\lambda$ on the central subgroup $V \cong \cyclicg{2}$.
\begin{align}
	\rho_0(\elembigg{g}) = \left\{ \begin{array}{l l}
		1 & \text{if } \elembigg{g}=(1_\biggroup,E)\\
		-1 & \text{if } \elembigg{g}=(1_\biggroup,F)
	\end{array} \right.
\end{align}

Now, the central subgroup $V$ is also the commutator subgroup $\commutatorsg{X}$ of $X$. This can be seen by noting that  $\biggroup = X/V$ is abelian. Thus, any commutator $x y \inv{x} \inv{y} \in V $, where $x \in X, y \in X$. If $x,y$ do not commute, then $x y \inv{x} \inv{y} = (1_\biggroup, F)$ and $\rho_0(x) \rho_0(y) \inv{\rho_0(x)} \inv{\rho_0(y)} = -1$. If $x \notin \gcenter{X}$ (center of $X$) then $\rho_0(x) = - \rho_0(y) \rho_0(x) \inv{\rho_0(y)}$ for some $y \in X$. Hence $\tr{\rho_0(x)} = 0$ $\forall x \notin \gcenter{X}$. Thus, we have
\begin{align}
	\chi_{\rho_0} (x) =\tr{\rho_0(x)} = \left\{ \begin{array}{l l} \gamma(x) d_{\rho_0}  & \text{if } x \in \gcenter{X} \\ 0 & \text{if } x \notin \gcenter{X}
	\end{array} \right.
\end{align}
where $\gamma$ is a one-dimensional complex representation of the center $\gcenter{X}$, and $d_{\rho_0}$ is the dimension of the irreuducible representation $\rho_0$. i.e., $|\chi_{\rho_0}(x)|= d_{\rho_0}$ if $x \in \gcenter{X}$, $0$ otherwise. Using the relation $\sum_{x \in X} |\chi_{\rho_0}(x)|^2 = |X|$, 
$
	d_{\rho_0}= \sqrt{\frac{|X|}{|\gcenter{X}|}}
$.
Let $X=(\elembigg{g},a)$. Then
$
	x \in \gcenter{X} \iff x y =y x \, \forall y \in X \iff \alpha(\elembigg{g}, \elembigg{h}) = \alpha(\elembigg{h}, \elembigg{g}) \, \forall  \elembigg{h} \in \biggroup 
	\iff \Omega(\elembigg{g}, \elembigg{h}) = \Omega(\elembigg{h}, \elembigg{g}) \, \forall \elembigg{h} \in \biggroup \iff U(\elembigg{g})  U(\elembigg{h}) = U(\elembigg{h})  U(\elembigg{g}) \forall \elembigg{h} \in \biggroup
$.
Define $ \mathcal{M} = \{ \elembigg{g} \in \biggroup | U(\elembigg{g}) U(\elembigg{h}) = U(\elembigg{h}) U(\elembigg{g}) \forall h \in \biggroup \} < \biggroup$, a subgroup of $\biggroup$. Then $\gcenter{X} = \{ (\elembigg{g},a) | \elembigg{g} \in \mathcal{M}, a \in \cyclicg{2}\}$, $|\gcenter{X}| = 2\times |\mathcal{M}|$. Hence the dimension of any irrep is
\begin{align}\label{eqn:degen}
	d_{\rho_0} = \sqrt{\frac{|\biggroup|}{|{\cal M}| }}
\end{align}

\end{document}